\documentclass{nature}
\usepackage{graphicx}

\usepackage[utf8]{inputenc}
\usepackage{amssymb} 
\usepackage{graphicx}
\usepackage{epstopdf}
\usepackage{multirow}
\usepackage{gensymb}
\usepackage{subcaption}
\usepackage{float}
\usepackage{natbib}
\usepackage[symbol]{footmisc}

\usepackage{color}
\usepackage[usenames,dvipsnames,table]{xcolor}
\title{Gamma-ray echoes from SS~433}
\author{\Large{Pol Bordas$^{1}$}}

\def\SS~433{{\em SS~433}}

\makeatletter
\let\saved@includegraphics\includegraphics
\AtBeginDocument{\let\includegraphics\saved@includegraphics}
\renewenvironment*{figure}{\@float{figure}}{\end@float}
\makeatother

\begin{document}
\maketitle

\begin{affiliations}

\item Departament F\'{i}sica Qu\`{a}ntica i Astrof\'{i}sica, Universitat de Barcelona. Institut de Ci\`{e}ncies del Cosmos (ICCUB). Institut de Ci\`{e}ncies Espacials de Catalunya (IEEC-CSIC). Mart\'{i} i Franqu\`{e}s 1, 08028, Barcelona, Spain.
%\item Departament F\'{i}sica Qu\`{a}ntica i Astrof\'{i}­sica, Universitat de Barcelona. Institut de Ci\`{e}ncies del Cosmos (ICCUB). Institut de Ci\`{e}ncies Espacials de Catalunya (IEEC-CSIC). Mart\'{i}­ i Franqu\`{e}s 1, 08028, Barcelona, Spain.
\end{affiliations}

\begin{abstract}

The detection of two sources of gamma rays towards the microquasar SS~433 has been recently reported \citep{Li2020}. The first source can be associated with SS~433's eastern jet lobe, whereas the second source is variable and displays significant periodicity compatible with the precession period of the binary system, of about 160 days. The location of this variable component is not compatible with the location of SS~433 jets. To explain the observed phenomenology, a scenario based on the illumination of dense gas clouds by relativistic protons accelerated at the interface of the accretion disk envelope has been proposed. Energetic arguments strongly constrain this scenario, however, as it requires an unknown mechanism capable to periodically channel a large fraction of SS~433's kinetic energy towards an emitter located $\sim 36$ parsec away from the central binary system. 

\end{abstract}

Only two X-ray binary systems (XRBs) displaying relativistic jets, Cyg X-1 and Cyg X-3, have so far been confirmed as high-energy gamma-ray emitters \citep{Tavani2009, Zanin2016}. Gamma rays are thought to be produced close to the jet base, where the ambient matter and radiation fields are the highest and processes like inverse Compton and proton–proton interactions can efficiently produce the observed fluxes. Recently, \citet{Li2020} have reported on the surprising discovery that a third accreting XRB, SS~433, is also producing gamma rays, but the emission is in this case resolved into several components, one of which displays significant periodicity and is located 36 parsecs away from the central binary system.

The micro-quasar SS~433 is beyond doubt a flagship of XRBs \citep{Stephenson_and_Sanduleak1977}. It is composed of a compact object, most probably a black hole of 10--20 solar masses (M$_{\odot}$), orbiting a 20M$_{\odot}$ supergiant star every P$_{\rm orb} \approx$ 13 days. The supergiant feeds the compact object at a rate of 10$^{-4}$ M$_{\odot}$ per year through the formation of an extended accretion disk that precesses under the pull of the secondary star every P$_{prec} \approx$ 162 days. SS~433 stands as the only supercritical accretor known in our Galaxy \citep{Fabrika2004}; it has an unparalleled energy budget delivering more than a million times the Sun's luminosity, and is identified as a true cosmic-ray factory and the sole case in which protons are detected in relativistic jets \cite{Migliari2002}. Since its discovery, SS~433 has been a unique laboratory for high-energy astrophysics, shaping our knowledge on XRBs and the mechanism through which accretion-powered jets are produced.
The first claims for the detection of gamma-ray activity towards SS~433 came in 2015 after the analysis of about five years of data taken with the Fermi Large Area Telescope (LAT) satellite \citep{Bordas2015}. Gamma-ray emission with a spectral energy distribution peaking at a few hundred MeV was detected from a region encompassing SS~433 and the surrounding W50 nebula. No signs of flux variability were seen, suggesting that gamma rays were produced at the SS~433/W50 interaction regions. Follow-up studies at high energies found a tentative extended nature for the gamma-ray source \citep{Sun2019} and also provided some hints of variability compatible with the precession period of the accretion disk \citep{Rasul2019}. Monitoring observations conducted with the High Altitude Water Cherenkov array resulted in the detection of gamma rays from the east and west lobes of SS~433/W50 at energies above 25 TeV \citep{HAWC2018}. This is the first and so far only confirmed case of gamma-ray emission detected from jet-inflated bubbles in our Galaxy.

Using a larger Fermi LAT data set and the latest source catalogues and models for the Galactic diffuse emission, \citet{Li2020} consider in detail the strong contamination from the bright and nearby pulsar PSR J1907+0602, a key factor in reducing potentially severe spectral and morphological systematic effects. Employing the latest ephemeris for this pulsar, each recovered gamma ray is imprinted with information on PSR J1907+0602's rotational phase. Selecting only off-peak intervals in their data set, the results of this ``gating'' analysis have revealed that gamma rays are produced relatively far from the position of the central binary system and are composed of several components. On the one hand, steady gamma ray emission is recovered from a source spatially coincident with the western terminal jet lobe. On the other hand, a new variable component has emerged, named Fermi J1913+0515 (see Fig. 1), displaying a periodic modulation with a statistical significance of 3.5$\sigma$ at a period of $\sim$160 days. This periodicity is compatible with the precession period of SS~433, but with an important caveat: the emitter seems to be located more than 35 parsecs away from it.
Theoretical predictions of variable gamma-ray emission based on the periodic changes of the conditions of the medium around the central system \citep{Reynoso2008} therefore seem to be ruled out. Li and collaborators report, however, on the presence of dense clouds of atomic gas, with a mass of about 25,000 M$_\odot$, in SS~433's surroundings that are spatially coincident with this variable gamma-ray component. Gamma rays could therefore have a hadronic origin, following the interaction of relativistic protons flowing along SS~433 jets against target material from the clouds \citep{Bosch-Ramon2005}. But once more, gamma rays cannot be the product of the illumination of these clouds by the jets, since these propagate along directions incompatible with the location of the gamma-ray excess. Intriguingly, and against all expectations, the observed variable gamma-ray component seems not be related to SS~433 jets.
The supercritical accretion regime operating in SS~433 not only creates powerful jets, though, it also releases large amounts of kinetic energy through the creation of equatorial relativistic outflows or winds produced at the surface layers of the accretion disk \citep{Sadowski2014}. Such a high kinetic power and the geometric inclination of the disk-wind envelope and its relatively wide solid angle, amplified by precession, prompted \citet{Li2020} to suggest this wind as an additional power source from SS~433 that could be illuminating the gas clouds. In order to keep periodicity, however, relativistic protons should regularly arrive at the cloud at a sufficient rate.

For a gamma-ray luminosity at a level of $L_{\gamma} = 10^{34}$~erg/s, and a cloud density $n = 20$~cm$^{-3}$, an energy reservoir in protons of $W_{\rm p} = 2.5 \times 10^{48}$ ($L_{\gamma}/10^{34}$)($n$/20) erg is needed. The kinetic power in the wind is of the order of $L_{\rm kin} \approx 10^{39}$~erg/s, so that about 100 yrs of continuous injection would be needed. Such a relatively large time would make however that the $\sim 160$d periodicity would be lost. The only way to circumvent this problem, both from an energetics basis and to explain the observed phenomenology, is to assume a local enhancement of the cloud density (a cusp with $n \gtrsim 10^{4}$~cm$^{-3}$), and to postulate a mechanism capable of delivering periodically about 10\% of the whole SS~433 kinetic power in relativistic protons for about half of its precession cycle ($\sim 80$ days) and place them right at the location of the cusp. On theoretical grounds this seems unachievable. Not only is a rather efficient acceleration mechanism needed to get $\sim 0.1$ L$_{\rm kin}$ in relativistic protons, one also needs to funnel all this power towards the cloud, located 36 parsecs away, through some unknown streaming mechanism that connects it with the azimuthally isotropic and hence rather uncollimated disk-wind envelope.

More than 40 years after its discovery, SS~433 keeps providing remarkable observational and theoretical challenges for the high-energy astrophysics community. The gamma-ray band wasn't going to be an exception, and the detection of emission from the jet lobes and the presence of a new periodic gamma-ray signal tens of parsecs away from the central system are beyond any previous expectation. Additional observations at this energy band, particularly with the forthcoming Cherenkov Telescope Array observatory, will provide further insights into the high-energy processes taking place in SS~433. In parallel, the development of a new framework providing theoretical support for efficient particle acceleration and anisotropic diffusion of relativistic winds from accretion disks is still to be forged out.

 \begin{figure}
 \centering
 \includegraphics[width=0.75\linewidth]{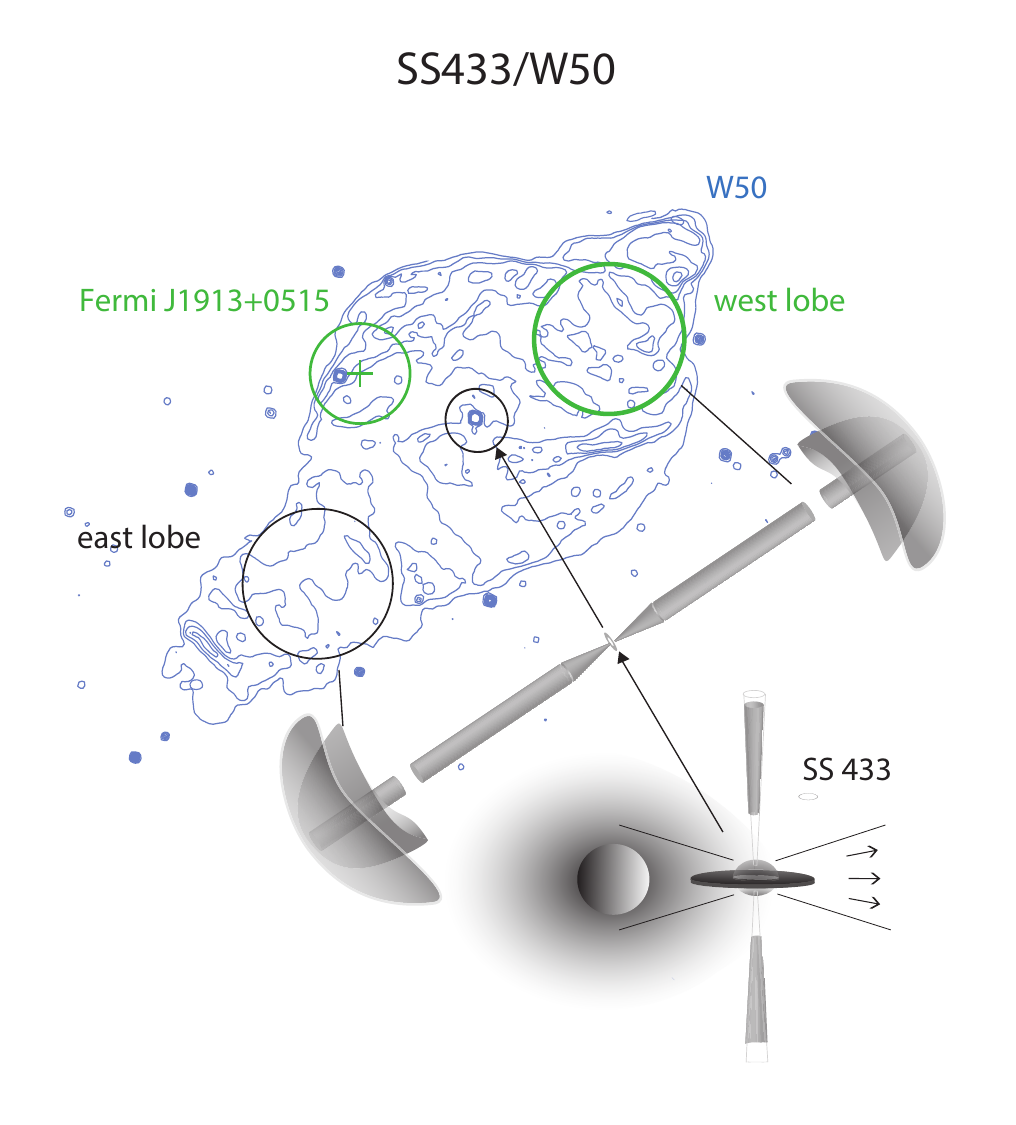}
 \caption{Figure 1: Sketch of the micro-quasar SS~433 and the W50 radio nebula seen in radio \citep{Dubner1998}. Black circles indicate the jet–medium interaction regions and green circles indicate the two components displaying gamma-ray emission as reported in \citet{Li2020}. Whereas gamma-rays from the west lobe display steady fluxes, periodic emission is detected from Fermi J1913+0515, located $\sim$36 parsecs away from SS~433.}
 \label{fig:stream}
 \end{figure}

\newpage

\noindent {\textbf{ACKNOWLEDGEMENTS}}

\small{ \noindent P. B. acknowledges the support by the Spanish Ministerio de Econom\'{i}a, Industria y Competitividad (MINEICO/FEDER, UE) under grant AYA2016-76012-C3-1-P, the Ministerio de Ciencia e Innovaci\'{o}n (MICINN) under grant PID2019-105510GB-C31 and through the ``Center of Excellence Mar\'{i}a de Maeztu 2020-2023'' award to the ICCUB (CEX2019-000918-M), and by the Catalan DEC grant 2017 SGR 643.}

\noindent {\textbf{REFERENCES}}

\bibliography{bibliography}

\begin{thebibliography}{14}
\expandafter\ifx\csname natexlab\endcsname\relax\def\natexlab#1{#1}\fi

\bibitem[{{Abeysekara} {et~al.}(2018){Abeysekara}, {Albert}, {Alfaro},
  {Alvarez}, {{\'A}lvarez}, {Arceo}, {Arteaga-Vel{\'a}zquez}, {Avila Rojas},
  {Ayala Solares}, {Belmont-Moreno}, {BenZvi}, {Brisbois}, {Caballero-Mora},
  {Capistr{\'a}n}, {Carrami{\~n}ana}, {Casanova}, {Castillo}, {Cotti},
  {Cotzomi}, {Couti{\~n}o de Le{\'o}n}, {De Le{\'o}n}, {De la Fuente},
  {D{\'\i}az-V{\'e}lez}, {Dichiara}, {Dingus}, {DuVernois}, {Ellsworth},
  {Engel}, {Espinoza}, {Fang}, {Fleischhack}, {Fraija}, {Galv{\'a}n-G{\'a}mez},
  {Garc{\'\i}a-Gonz{\'a}lez}, {Garfias}, {Gonz{\'a}lez-Mu{\~n}oz},
  {Gonz{\'a}lez}, {Goodman}, {Hampel-Arias}, {Harding}, {Hernandez}, {Hinton},
  {Hona}, {Hueyotl-Zahuantitla}, {Hui}, {H{\"u}ntemeyer}, {Iriarte},
  {Jardin-Blicq}, {Joshi}, {Kaufmann}, {Kar}, {Kunde}, {Lauer}, {Lee},
  {Le{\'o}n Vargas}, {Li}, {Linnemann}, {Longinotti}, {Luis-Raya},
  {L{\'o}pez-Coto}, {Malone}, {Marinelli}, {Martinez}, {Martinez-Castellanos},
  {Mart{\'\i}nez-Castro}, {Matthews}, {Mirand a-Romagnoli}, {Moreno},
  {Mostaf{\'a}}, {Nayerhoda}, {Nellen}, {Newbold}, {Nisa}, {Noriega-Papaqui},
  {Pretz}, {P{\'e}rez-P{\'e}rez}, {Ren}, {Rho}, {Rivi{\`e}re},
  {Rosa-Gonz{\'a}lez}, {Rosenberg}, {Ruiz-Velasco}, {Salesa Greus}, {Sandoval},
  {Schneider}, {Schoorlemmer}, {Seglar Arroyo}, {Sinnis}, {Smith}, {Springer},
  {Surajbali}, {Taboada}, {Tibolla}, {Tollefson}, {Torres}, {Vianello},
  {Villase{\~n}or}, {Weisgarber}, {Werner}, {Westerhoff}, {Wood}, {Yapici},
  {Yodh}, {Zepeda}, {Zhang}, \& {Zhou}}]{HAWC2018}
{Abeysekara}, A.~U., {Albert}, A., {Alfaro}, R., {et~al.} 2018, Nature, 562, 82

\bibitem[{{Bordas} {et~al.}(2015){Bordas}, {Yang}, {Kafexhiu}, \&
  {Aharonian}}]{Bordas2015}
{Bordas}, P., {Yang}, R., {Kafexhiu}, E., \& {Aharonian}, F. 2015, ApJL, 807,
  L8

\bibitem[{{Bosch-Ramon} {et~al.}(2005){Bosch-Ramon}, {Aharonian}, \&
  {Paredes}}]{Bosch-Ramon2005}
{Bosch-Ramon}, V., {Aharonian}, F.~A., \& {Paredes}, J.~M. 2005, A\&A, 432, 609

\bibitem[{{Dubner} {et~al.}(1998){Dubner}, {Holdaway}, {Goss}, \&
  {Mirabel}}]{Dubner1998}
{Dubner}, G.~M., {Holdaway}, M., {Goss}, W.~M., \& {Mirabel}, I.~F. 1998, AJ,
  116, 1842

\bibitem[{{Fabrika}(2004)}]{Fabrika2004}
{Fabrika}, S. 2004, ApSPr, 12, 1

\bibitem[{{Li} {et~al.}(2020){Li}, {Torres}, {Liu}, {Kerr}, {de O{\~n}a
  Wilhelmi}, \& {Su}}]{Li2020}
{Li}, J., {Torres}, D.~F., {Liu}, R.-Y., {et~al.} 2020, Nature Astronomy

\bibitem[{{Migliari} {et~al.}(2002){Migliari}, {Fender}, \&
  {M{\'e}ndez}}]{Migliari2002}
{Migliari}, S., {Fender}, R., \& {M{\'e}ndez}, M. 2002, Science, 297, 1673

\bibitem[{{Rasul} {et~al.}(2019){Rasul}, {Chadwick}, {Graham}, \&
  {Brown}}]{Rasul2019}
{Rasul}, K., {Chadwick}, P.~M., {Graham}, J.~A., \& {Brown}, A.~M. 2019, MNRAS,
  485, 2970

\bibitem[{{Reynoso} {et~al.}(2008){Reynoso}, {Romero}, \&
  {Christiansen}}]{Reynoso2008}
{Reynoso}, M.~M., {Romero}, G.~E., \& {Christiansen}, H.~R. 2008, MNRAS, 387,
  1745

\bibitem[{{Sadowski} {et~al.}(2014){Sadowski}, {Narayan}, {McKinney}, \&
  {Tchekhovskoy}}]{Sadowski2014}
{Sadowski}, A., {Narayan}, R., {McKinney}, J.~C., \& {Tchekhovskoy}, A. 2014,
  MNRAS, 439, 503

\bibitem[{{Stephenson} \& {Sanduleak}(1977)}]{Stephenson_and_Sanduleak1977}
{Stephenson}, C.~B. \& {Sanduleak}, N. 1977, ApJs, 33, 459

\bibitem[{{Sun} {et~al.}(2019){Sun}, {Yang}, {Liu}, {Xi}, \& {Wang}}]{Sun2019}
{Sun}, X.-N., {Yang}, R.-Z., {Liu}, B., {Xi}, S.-Q., \& {Wang}, X.-Y. 2019,
  A\&A, 626, A113

\bibitem[{{Tavani} {et~al.}(2009){Tavani}, {Bulgarelli}, {Piano}, {Sabatini},
  {Striani}, {Evangelista}, {Trois}, {Pooley}, {Trushkin}, {Nizhelskij},
  {McCollough}, {Koljonen}, {Pucella}, {Giuliani}, {Chen}, {Costa},
  {Vittorini}, {Trifoglio}, {Gianotti}, {Argan}, {Barbiellini}, {Caraveo},
  {Cattaneo}, {Cocco}, {Contessi}, {D'Ammand o}, {Del Monte}, {de Paris}, {Di
  Cocco}, {di Persio}, {Donnarumma}, {Feroci}, {Ferrari}, {Fuschino}, {Galli},
  {Labanti}, {Lapshov}, {Lazzarotto}, {Lipari}, {Longo}, {Mattaini},
  {Marisaldi}, {Mastropietro}, {Mauri}, {Mereghetti}, {Morelli}, {Morselli},
  {Pacciani}, {Pellizzoni}, {Perotti}, {Picozza}, {Pilia}, {Prest},
  {Rapisarda}, {Rappoldi}, {Rossi}, {Rubini}, {Scalise}, {Soffitta},
  {Vallazza}, {Vercellone}, {Zambra}, {Zanello}, {Pittori}, {Verrecchia},
  {Giommi}, {Colafrancesco}, {Santolamazza}, {Antonelli}, \&
  {Salotti}}]{Tavani2009}
{Tavani}, M., {Bulgarelli}, A., {Piano}, G., {et~al.} 2009, Nature, 462, 620

\bibitem[{{Zanin} {et~al.}(2016){Zanin}, {Fern{\'a}ndez-Barral}, {de O{\~n}a
  Wilhelmi}, {Aharonian}, {Blanch}, {Bosch-Ramon}, \& {Galindo}}]{Zanin2016}
{Zanin}, R., {Fern{\'a}ndez-Barral}, A., {de O{\~n}a Wilhelmi}, E., {et~al.}
  2016, A\&A, 596, A55

\end{thebibliography}

% \begin{thebibliography}{widest entry}
% \bibitem[label1]{cite_key1} bibliographic information \bibitem[label2]{cite_key2} bibliographic information
% \end{thebibliography}

\end{document}